\documentclass[12pt]{article}

\usepackage[a4paper,left=30mm,right=20mm,top=20mm,bottom=20mm]{geometry}
\usepackage{graphics}
\usepackage{amsmath}
\usepackage{amssymb}
\usepackage{epsfig}
\usepackage{cite}
\usepackage{xcolor}
\usepackage[unicode]{hyperref}
\newcommand{\be}{\begin{eqnarray}}
\newcommand{\ee}{\end{eqnarray}}
\newcommand{\beq}{\begin{equation}}
\newcommand{\eeq}{\end{equation}}

\title{Self-consistent description of HERA data at low $Q^2$ and soft hadron production at LHC}

\author{A.V.~Lipatov$^{1,2}$, G.I.~Lykasov$^{2}$, M.A.~Malyshev$^{1,2}$}

\begin{document}

\maketitle

\begin{center}

{\it $^{1}$Skobeltsyn Institute of Nuclear Physics, Lomonosov Moscow State University, 119991, Moscow, Russia}\\
{\it $^{2}$Joint Institute for Nuclear Research, 141980, Dubna, Moscow region, Russia}\\

\end{center}

\vspace{0.5cm}

\begin{center}

{\bf Abstract }

\end{center}

\indent 

Using the analytical expression for transverse momentum dependent (TMD) gluon density
in a proton, a self-consistent simultaneous description of low $Q^2$ data on
%$\gamma^* p$
proton structure function $F_2(x,Q^2)$,
reduced cross section for the electron-proton deep inelastic scattering at HERA and soft hadron production in $pp$ collisions at
the LHC is achieved in the framework of color dipole approach and modified
quark-gluon string model.

%Using the starting gluon TMD  
%the self-consistent description of the $ep$ DIS at low $x$ and small
%$Q^2$ and the soft hadron production in $pp$ collisions at LHC energies in the
%mid-rapidity region is presented. 

\vspace{1cm}


\newpage

It is well known that the deep inelatic electron-proton
  scattering (DIS) processes
at low $Q^2$ and small Bjorken variable $x$ can provide important
information about the non-perturbative parton (quark and gluon) structure of the
proton.
Detailed knowledge of the latter is necessary for any theoretical study of different
high energy processes performed within the Quantum Chromodynamics (QCD).
Moreover, it is necessary for future experiments planned at
the Large Hadron electron Collider (LHeC)\cite{LHeC} and Future Circular
hadron-electron Collider
(FCC-he)\cite{FCC}, where facilities for using electron-proton center-of-mass
energies $\sqrt s = 1.3$~TeV
and $\sqrt s = 3.5$~TeV are proposed. 

In the region of relatively low $Q^2 \lesssim 4$~GeV$^2$ and
  small $x$ the conventional 
  (collinear) QCD approach based on Dokshitzer-Gribov-Lipatov-Altarelli-Parisi
  (DGLAP) evolution
equations\cite{DGLAP} has some difficulties to describe the DIS experimental data.
For example, next-to-leading order (NLO) QCD corrections to the proton longitudinal
structure function $F_L(x, Q^2)$ are large and negative, that could lead to the
negative $F_L(x, Q^2)$ values\cite{FL-NLO-Thorne-1, FL-NLO-Thorne-2}.
However, in the considered kinematical regime the perturbative expansion contains
also large logarithmic 
terms proportional to $\alpha_s^n \ln^n 1/x$.
Resummation of these terms, in general, is necessary to produce realiable
theoretical predictions. % which are in agreement with the collider data.
Such resummation can be done in the framework of high-energy
\cite{HighEnergyFactorization} or $k_T$-factorization\cite{kt-factorization}
approach with 
using a transverse momentum dependent (TMD, or unintegrated)
gluon density $f_g(x,{\mathbf k}_T^2,\mu^2)$ obeying the Balitsky-Fadin-Kuraev-Lipatov (BFKL)\cite{BFKL} or
Ciafaloni-Catani-Fiorani-Marchesini (CCFM)\cite{CCFM} evolution
equations.

%%%%%%%%%%%%%%%%%%%%%%%%%%%%%%%%%%%%%%%%%%%%%%%%%%%%%%%%%%%%%%%%%%%%%%%%%%%%%

As it is suggested in \cite{kt-factorization}, at very low $x$ and $Q^2$
the proton structure function $F_2(x,Q^2)$ 
%determined mainly by the TMD gluon density $f_g(x,k_T,Q^2)$ 
can be saturated (does not depend on $Q^2$). In\cite{Saturation-Mueller}
the partial saturation of the TMD gluon density at $Q^2$ less than the saturation scale $Q_s^2$ was 
suggested: $f_g(x,{\mathbf k}_T^2,\mu^2) \simeq const \times \ln 1/x$.
Further developments on the saturation effect were done\cite{DP-NZ-1, DP-NZ-2, GBW1, GBW2}.
%%%%%%%%%%%%%%%%%%%%%%%%%%%%%%%%%%%%%%%%%%%%%%%%%%%%%%%%%%%%%%%%%%%%%%%%%%%%%
At asympotically low $x$, an equivalent description of early HERA data on $F_2(x,Q^2)$ was
  provided by the color dipole model \cite{DP-NZ-1, DP-NZ-2, GBW1, GBW2},
  where gluon saturation effects \cite{Saturation-Mueller} important at low scales
  (and, of course, at low $Q^2$ in DIS) can be consistently taken into account.
In this approach, the DIS events are
considered as an interaction of the virtual photon decaying into a color dipole
$q{\bar q}$ and a proton.
The saturation of the photon-proton cross section $\sigma^{\gamma^* p}(x, Q^2)$ as
a function of the transverse 
distance $r$ between $q$ and ${\bar q}$ is suggested at large $r$ or small $Q^2$
(see\cite{GBW1, GBW2} for more information).

In our previous study\cite{Input-1} the color dipole model
was used to investigate the connection between the DIS processes studied at HERA
and soft hadron production at LHC\footnote{Within the modified quark-gluon 
string model (QGSM)
\cite{ModifiedSoftQuarkGluonStringModel-1, ModifiedSoftQuarkGluonStringModel-2}
(see also\cite{SoftQuarkGluonStringModel-1, SoftQuarkGluonStringModel-2}).}.
An analytical expression for the TMD gluon 
density in a proton at the low scale $\mu_0$ (which is of order of the hadronic scale,
$\mu_0 \sim 1$~GeV) was proposed
and its saturation dynamics was discussed.
Further improvements of such investigations were developed\cite{Input-2, Input-3, Input-4, LLM-2022}.
In particular, it was shown that a number of hard LHC processes 
can be reasonably well described
within the $k_T$-factorization framework
if the proposed TMD gluon distribution is considered
as the input for subsequent (non-collinear) CCFM evolution
(see, for example,\cite{LLM-2022} and references therein).
In the present note we continue our study 
and concentrate on self-consistent 
simultaneous description of latest HERA data
on the reduced cross section $\sigma_r(x, Q^2)$ at low $Q^2$\cite{sigma_red-ZEUS+H1} (and, consequently, $F_2(x,Q^2)$ data\cite{sigma-gamma-proton-ZEUS, sigma-gamma-proton-H1})
and LHC data on charged hadron production at small transverse momenta $p_T$ in
the mid-rapidity region\cite{SoftHadronData-1,SoftHadronData-2,SoftHadronData-3}.
Note that latest HERA data\cite{sigma_red-ZEUS+H1,sigma-gamma-proton-ZEUS, sigma-gamma-proton-H1} have not been considered yet
within the developed approach.

%In \cite{GLLZ:2013} the connection of the DIS at low $Q^2$ with the soft hadron
%production in $pp$ collisions at LHC energies in the mid-rapidity region and
%%\cite{BGLP:2012} and \cite{LLM:2023}, \cite{LLM2:2023} 
%small transverse momenta of hadrons was investigated.
%Using the dipole model \cite{GBW1,GBW2} and calculating the gluon density
%at low $Q^2$ \cite{BGLP:2012} entering into the CCFM evolution equation
%\cite{CCFM}, as a starting gluon function, many observables of hard $pp$
%processes
%and the soft hadron production at LHC energies were satisfactorily described.
%The further improvement of such investigation was developed in
%\cite{GLLZ:2013,BGLP:2012,LLZ:2014,GLLZ:2016,Abdulov:2018,LLM:2023,LLM2:2023}.  
%However, the starting TMD gluon distribution $f^{(0)}_g(x,k_T$ obtained in
%\cite{LLM:2023} does not allow us to describe the proton structure function
%$F_2(x,Q^2)$ and the effective cross section of the $\gamma^* p$ interaction.
%In this paper we perform the self-consistent calculation of these DIS
%observables and inclusive hadron spectra in $pp$ collisions at LHC energies,
%and low transverse momenta $p_T$ in the mid-rapidity region.

%%%%%%%%%%%%%%%%%%%%%%%%%%%%%%%%%%%%%%%%%%%%%%%%%%%%%%%%%%%%%%%%%%%%%%%%%%
%%%%%%%%%%%%%%%%%%%%%%%%%%%%%%%%%%%%%%%%%%%%%%%%%%%%%%%%%%%%%%%%%%%%%%%%%%%
%%%%%%%%%%%%%%%%%%%%%%%%%%%%%%%%%%%%%%%%%%%%%%%%%%%%%%%%%%%%%%%%%%%%%%%%%%%
%\section{Saturation dynamics} \indent

For the reader's convenience, we recall some important formulas
%The main idea of the saturation in deep inelastic scattering (DIS) is based
on transition from high to low $Q^2$
%observed in the total $\gamma^* p$ cross section, where $\gamma^*$ is the
%virtual exchange photon 
in DIS.
As it was mentioned above, dipole formalism for calculation of
the deep inelastic and related diffractive cross sections of $\gamma^* p$
scattering at small $x$ was 
developed\cite{DP-NZ-1, DP-NZ-2,GBW1,GBW2}.
For transversely ($T$) and longitudinally ($L$) polarized photons
the $\gamma^* p$
  cross section
  $\sigma^{\gamma^* p}(x, Q^2) = \sigma_T(x, Q^2) + \sigma_L(x, Q^2)$
  can be presented in the following form: 
\begin{gather}
  \sigma_{T,\,L}(x,Q^2) = \int d^2{\mathbf r} \int\limits_0^1 dz
  |{\Psi}_{T,\,L}(z,r)|^2 \,\hat{\sigma}(x,r^2) = \nonumber \\
  = 2\pi \int\limits_0^\infty rdr \int\limits_0^1 dz
  |{\Psi}_{T,\,L}(z,r)|^2 \, \hat{\sigma}(x,r^2),
  \label{def:GBW_CrossSections}
\end{gather}
\noindent
where $z$ is the quark longitudinal momentum fraction ($1 - z$
  for antiquark) with respect to photon momentum $q$,
  $x = Q^2/(W^2 + Q^2)$, $Q^2 = - q^2$, $W^2 = (p + q)^2$ with $p$ being the proton
  momentum.
The squared photon wave functions read
\begin{gather}
  |{\Psi}_{T}(z,r)|^2 = \frac{6\alpha_{em}}{4\pi^2}\sum_f e_f^2
  \left([z^2 + (1-z)^2]\,\epsilon^2 K_1^2(\epsilon r) + m_f^2K_0^2(\epsilon r)
  \right), \nonumber \\
  |{\Psi}_{L}(z,r)|^2 = \frac{6\alpha_{em}}{4\pi^2}\sum_f e_f^2\left(4Q^2z^2(1-z)^2)
  K_0^2(\epsilon r)\right),
\label{def:GBW_Psi}
\end{gather}
\noindent
where $\epsilon^2 = z(1-z)Q^2 + m_f^2$, $K_0$ and $K_1$ are
McDonald functions and the summation is performed over the quark flavors $f$.
%%%%%%%%%%%%%%%%%%%%%%%%%%%%%%%%%%%%%%%%%%%%%%%%%%%%%%%%%%%%%%%%%%%%%%%%%%
%\section{Structure functions $F_t,F_l$ and cross section $\sigma_{\gamma* p}$}
%The structure function $F_2$ is the sum of the transverse structure
%function $F_T$ and longitudinal one $F_L$, i.e.,
%\begin{equation}
%F_2(x,Q^2)~=~F_T(x,Q^2)~+~F_L(x,Q^2) ~,
%\label{def:F2}
%\end{equation}
%Then, for $F_T$ and $F_L$ we have the following forms valid at low
% $x$ neglecting small term $m x^2$:
%\begin{equation}
%F_{T,L}(x,Q^2)~=~\frac{Q^2}{4\pi^2\alpha_{em}}\sigma_{T,L}(x,Q^2) ~,
%\label{def:FTL}
%\end{equation}
%%%%%%%%%%%%%%%%%%%%%%%%%%%%%%%%%%%%%%%%%%%%%%%%%%%%%%%%%%%%%%%%%%%%%
As it was originally assumed\cite{GBW1, GBW2}, the effective
  dipole cross section 
%as a function of the distance $r$ between $q$ and $\bar q$ 
is saturated at large $r$ and
presented in the following (GBW) form:
\begin{gather}
  \hat{\sigma}(x,r^2) = \sigma_0\left\{1-\exp\left(-\frac{r^2}{4R_0^2(x)}\right)
  \right\}, \quad R_0^2(x) = {1\over Q_0^2}\left(x\over x_0\right)^\lambda,
  \label{def:GBW_crsect}
\end{gather}
\noindent
where overall normalization $\sigma_0 = 29.12$~mb and parameters
  $\lambda = 0.277$, $x_0 = 4.1 \cdot 10^{-5}$, $Q_0 = 1$~GeV were found from
a fit to the inclusive DIS data. The relation of the TMD
gluon density in a proton $f_g(x, {\mathbf k}_T^2)$ to the
dipole cross section
$\hat{\sigma}(x,r^2)$ was calculated\cite{GBW2} within the two gluon
exchange approximation between color dipole $q{\bar q}$
  and proton debris. It has the
following form:
\begin{gather}
  \hat{\sigma}(x,r^2) = \frac{4\pi^2 \alpha_s}{3}\int\frac{d{\mathbf k}_T^2}
      {{\mathbf k}_T^2}\left\{1-J_0(|{\mathbf k}_T| r)\right\}
      f_g(x, {\mathbf k}_T^2),
  \label{def:GBW_dpcrsect}
\end{gather}
\noindent
where $J_0$ is the Bessel function of zero order and
  $\alpha_s = 0.2$. The relations (\ref{def:GBW_crsect}) and
  (\ref{def:GBW_dpcrsect})
lead to the GBW expression for the TMD gluon density in a proton:
\begin{gather}
  f_g(x, {\mathbf k}_T^2) = {3 \sigma_0\over 4 \pi^2 \alpha_s} R_0^2(x)
  {\mathbf k}_T^2 \exp\left( -R_0^2(x){\mathbf k}_T^2 \right). 
  \label{eq:GBW}
\end{gather}
\noindent
In our previous paper\cite{LLM-2022} another analytical form
  of the TMD gluon density
was proposed:
\begin{gather}
  f_{g}(x, {\mathbf k}_T^2) = c_g (1-x)^{b_g} \sum_{n = 1}^3 c_n \left[R_0(x)
    |{\mathbf k}_{T}|\right]^n \exp \left(-R_0(x)
  |{\mathbf k}_{T}|\right), \nonumber \\ 
  b_g = b_g(0) + {4C_A\over \beta_0}\ln{\alpha_s(Q_0^2)\over
    \alpha_s({\mathbf k}_{T}^2)},
  \label{eq:OurGluon}
\end{gather}
\noindent
where $C_A = N_C$, $\beta_0 = 11 - 2N_f/3$. All phenomenological parameters
essential at low $x$, namely, $c_g$, $c_1$, $c_2$ and $c_3$ were found from
the best description of 
recent LHC data\cite{SoftHadronData-1, SoftHadronData-2, SoftHadronData-3}
on charged soft hadron production at the mid-rapidity region
within the modified QGSM approach\cite{ModifiedSoftQuarkGluonStringModel-1,
  ModifiedSoftQuarkGluonStringModel-2}. 
Detailed information about the calculations and fitting procedure can be found in~\cite{LLM-2022}.
Note that we kept $x_0$ and $\lambda$ parameters as they were fitted in the
GBW model.
The TMD gluon density in a form (\ref{eq:OurGluon}), 
%\footnote{Referred as LLM'2022 set and available in the \textsc{tmdlib} package
%\cite{?}.,
being extended into a 
whole kinematical region by appying the subsequent CCFM evolution
and referred as the LLM'2022 set\footnote{Available now in the popular
\textsc{tmdlib} package\cite{TMDLib2}.},
is able to reproduce
the experimental data on number of processes studied at HERA and LHC colliders
(see also \cite{FL-our,Photon-our}).

With the listed pocket formulas we can now calculate some observables.
The proton structure function $F_2(x,Q^2)$ is the sum of the transverse structure
function $F_T(x, Q^2)$ and longitudinal one $F_L(x, Q^2)$:
\begin{equation}
F_2(x,Q^2) = F_T(x,Q^2) + F_L(x,Q^2),
\label{def:F2}
\end{equation}
\noindent
where $F_T(x, Q^2)$ and $F_L(x, Q^2)$ can be calculated at low
 $x$ neglecting small term $\sim m x^2$ using (\ref{def:GBW_CrossSections}):
\begin{equation}
F_{T,L}(x,Q^2) = \frac{Q^2}{4\pi^2\alpha_{em}}\sigma_{T,L}(x,Q^2),
\label{def:FTL}
\end{equation}
In fact, the reduced cross section $\sigma_r(x, Q^2)$ in the DIS %, which is related to
%struture functions $F_2$ and $F_L$, 
is measured experimentally. It can be calculated as:
\begin{equation}
  \sigma_r(x, Q^2)=\frac{Q^4x}{2\pi\alpha^2(1-y^2)}\frac{d^2\sigma}{dxdQ^2}=
  F_2(x,Q^2) - f(y)F_L(x,Q^2),
\label{def:sigmar}
\end{equation}
where $d^2\sigma/dxdQ^2$ is the double differential DIS cross section, $f(y)=y^2/(1+(1-y)^2)$, inelasticity $y=Q^2/(sx)$ and $s=4E_eE_p$ with $E_e$ and $E_p$ being the
electron and proton energies, respectively.

\begin{figure}
\begin{center}
  \includegraphics[width=15cm]{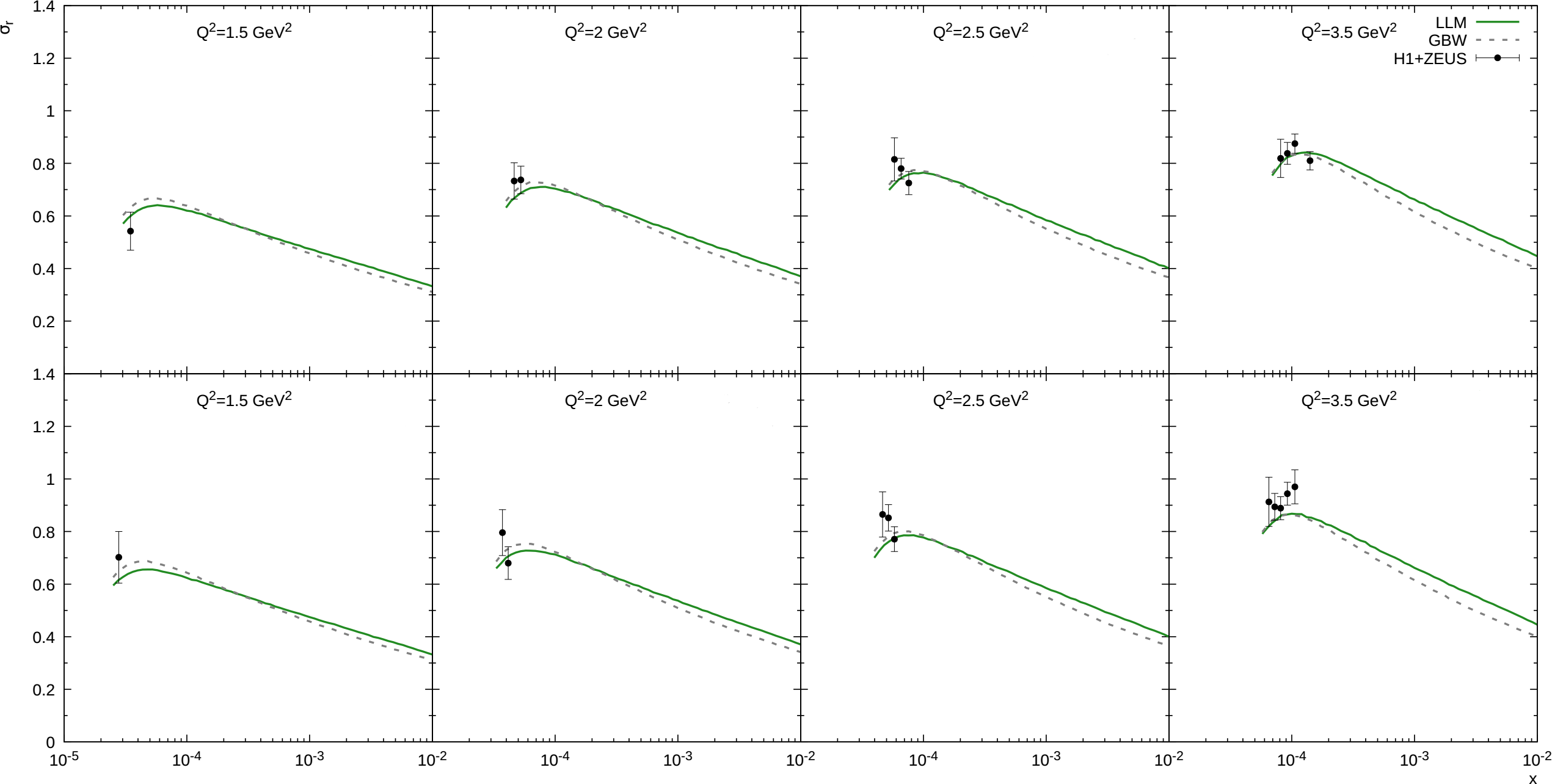}
  \caption{The reduced DIS cross section a function of $x$ at different $Q^2$ at $\sqrt{s}=225$~GeV (upper row) and $\sqrt{s}=251$~GeV (lower row). The solid green line corresponds to the results obtained with LLM TMD, and the dashed gray line shows GBW based results.
      The data are taken from ZEUS and H1~\cite{sigma_red-ZEUS+H1}.
}
\label{sigred1}
\end{center}
\end{figure}

\begin{figure}
\begin{center}
  \includegraphics[width=15cm]{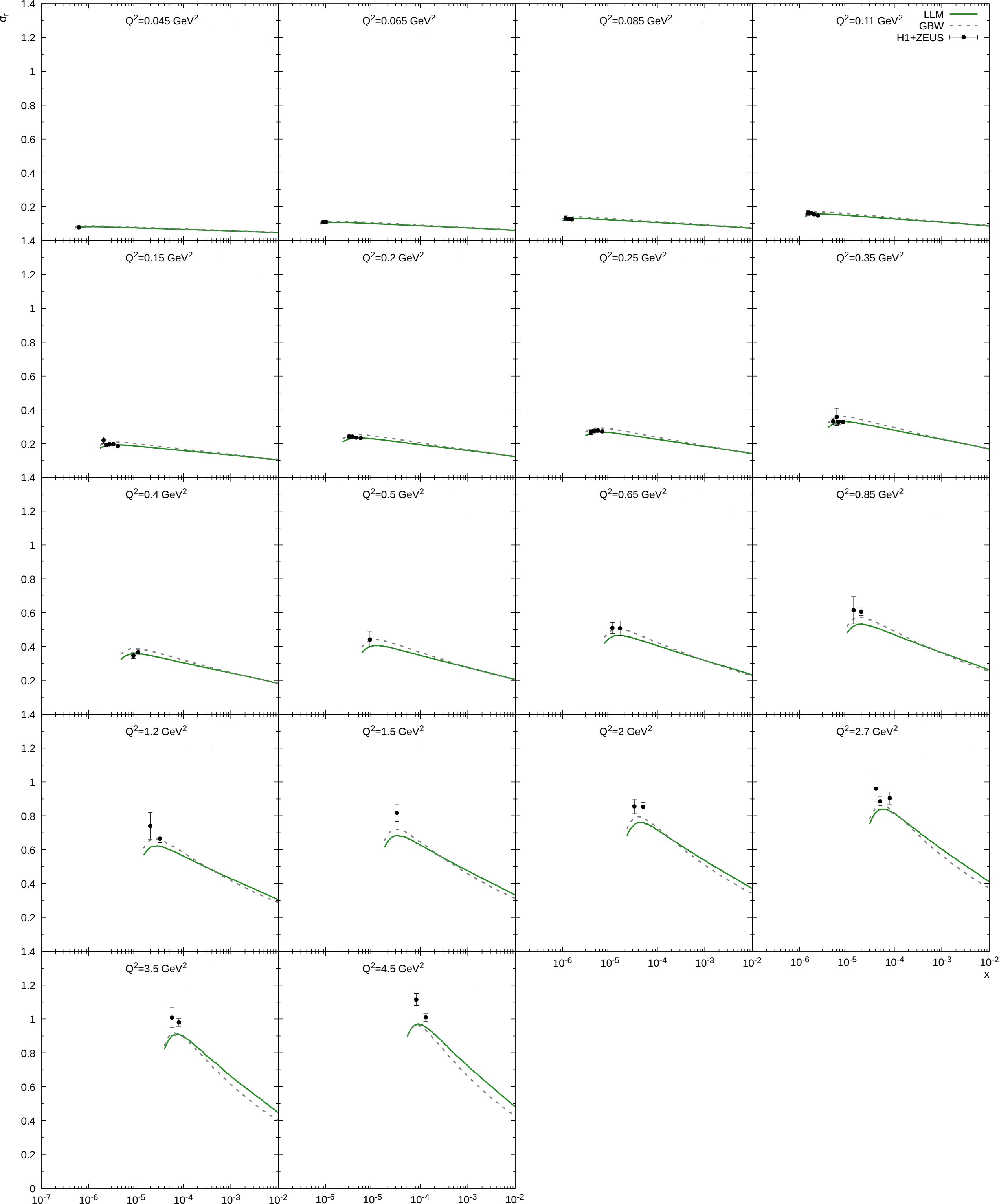}
  \caption{The reduced DIS cross section a function of $x$ at different $Q^2$ at $\sqrt{s}=300$~GeV. The notations are the same as on  Fig.~\ref{sigred1}.
      The data are taken from ZEUS and H1~\cite{sigma_red-ZEUS+H1}.
}
\label{sigred2}
\end{center}
\end{figure}

\begin{figure}
\begin{center}
  \includegraphics[width=15cm]{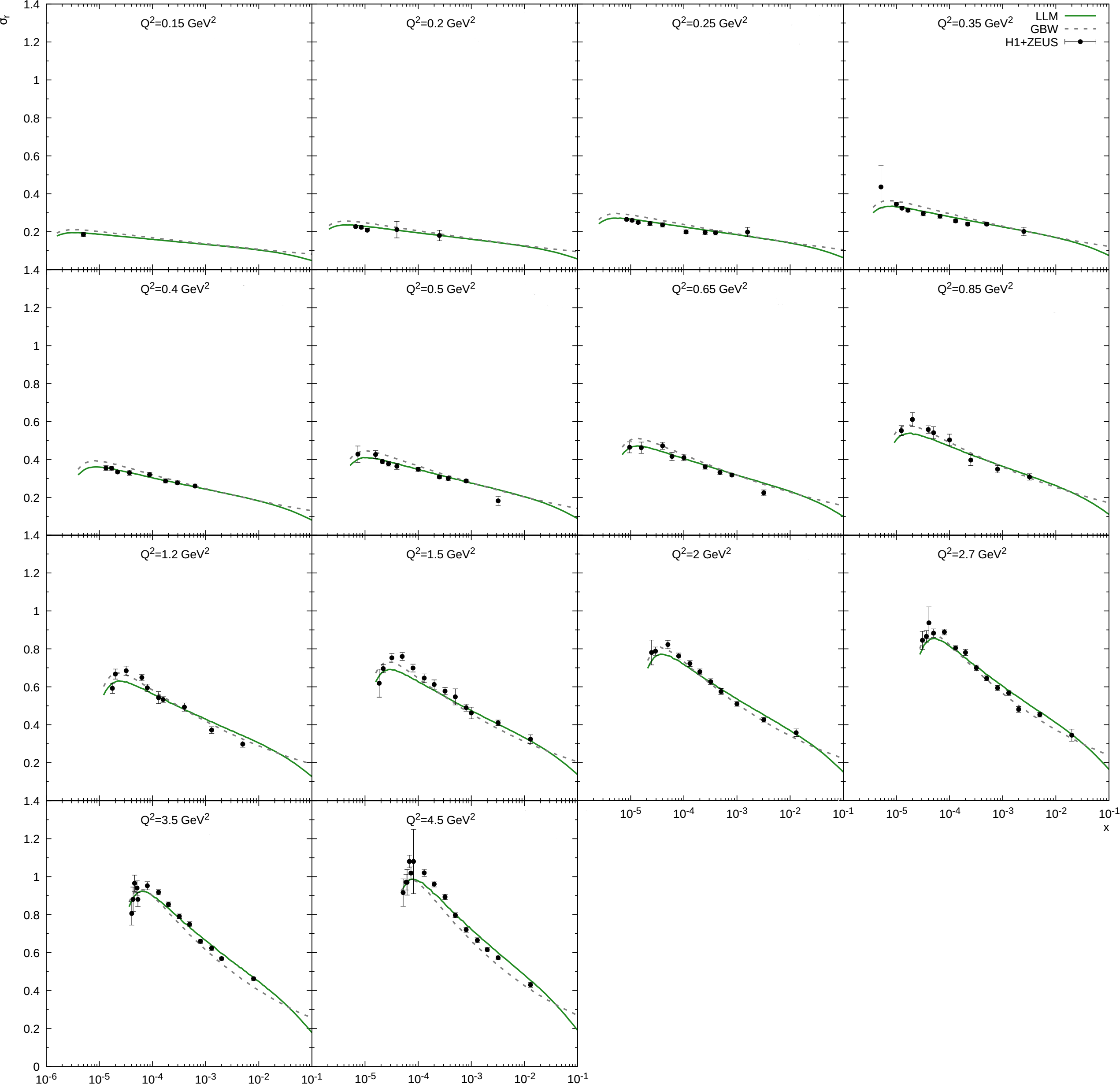}
  \caption{The reduced DIS cross section a function of $x$ at different $Q^2$ at $\sqrt{s}=318$~GeV. The notations are the same as on  Fig.~\ref{sigred1}.
      The data are taken from ZEUS and H1~\cite{sigma_red-ZEUS+H1}.
}
\label{sigred3}
\end{center}
\end{figure}

As it was mentioned above, here we try the LLM gluon to
describe the latest HERA data
\cite{sigma-gamma-proton-ZEUS, sigma-gamma-proton-H1}
on the reduced cross section $\sigma_r(x, Q^2)$ at low $Q^2$
within the dipole approach.
First of all, we find 
that we have to change the default values of $x_0$ and $\lambda$ parameters
involved in (\ref{def:GBW_crsect}) and (\ref{eq:OurGluon}).
Our calculations according to master formulas (\ref{def:GBW_CrossSections}) and
(\ref{def:GBW_Psi})
show that best description (with $\chi^2/n.d.f. = 2.3$) of the reduced cross section measured at different energies ($\sqrt{s}=225$, $251$, $300$ and $318$~GeV) 
is obtained with $x_0 = 1.3 \cdot 10^{-11}$
and $\lambda = 0.22$.
%Corresponding $\chi^2/n.d.f = ?$
%Some technical details of the calculations are collected in Appendix A.
Let us note that we analyzed the HERA data only in the low $Q^2$ region,
$Q^2 < 5$~GeV$^2$, 
since we set the lowest perturbative scale $Q_0 = 2.2$~GeV.
Above this scale, the effects of the QCD evolution can play a role.
Results of our fit are shown in Fig.~(\ref{sigred1})---(\ref{sigred3}), where
one can see the satisfactory description of the combined H1 and ZEUS data~\cite{sigma_red-ZEUS+H1}. 
The green line corresponds to our results obtained with newly fitted LLM formula (\ref{eq:OurGluon}) and 
gray dashed line shows results calculated within the GBW model taken with parameters 
determined\footnote{One can find another set of the GBW parameters in~\cite{Kutak1,Kutak2}, where Sudakov form-factor effects were taken into account.} by H1\cite{sigma-gamma-proton-H1}: $x_0=6.0\cdot 10^{-5}$ and $\lambda = 0.256$. 
We only increase the normalization by 10\%: $\sigma_0=27$~mb. One can see that GBW approach 
results in a worse description of the data ($\chi^2/n.d.f. = 4.1$). 

\begin{figure}
\begin{center}
  \includegraphics[width=15cm]{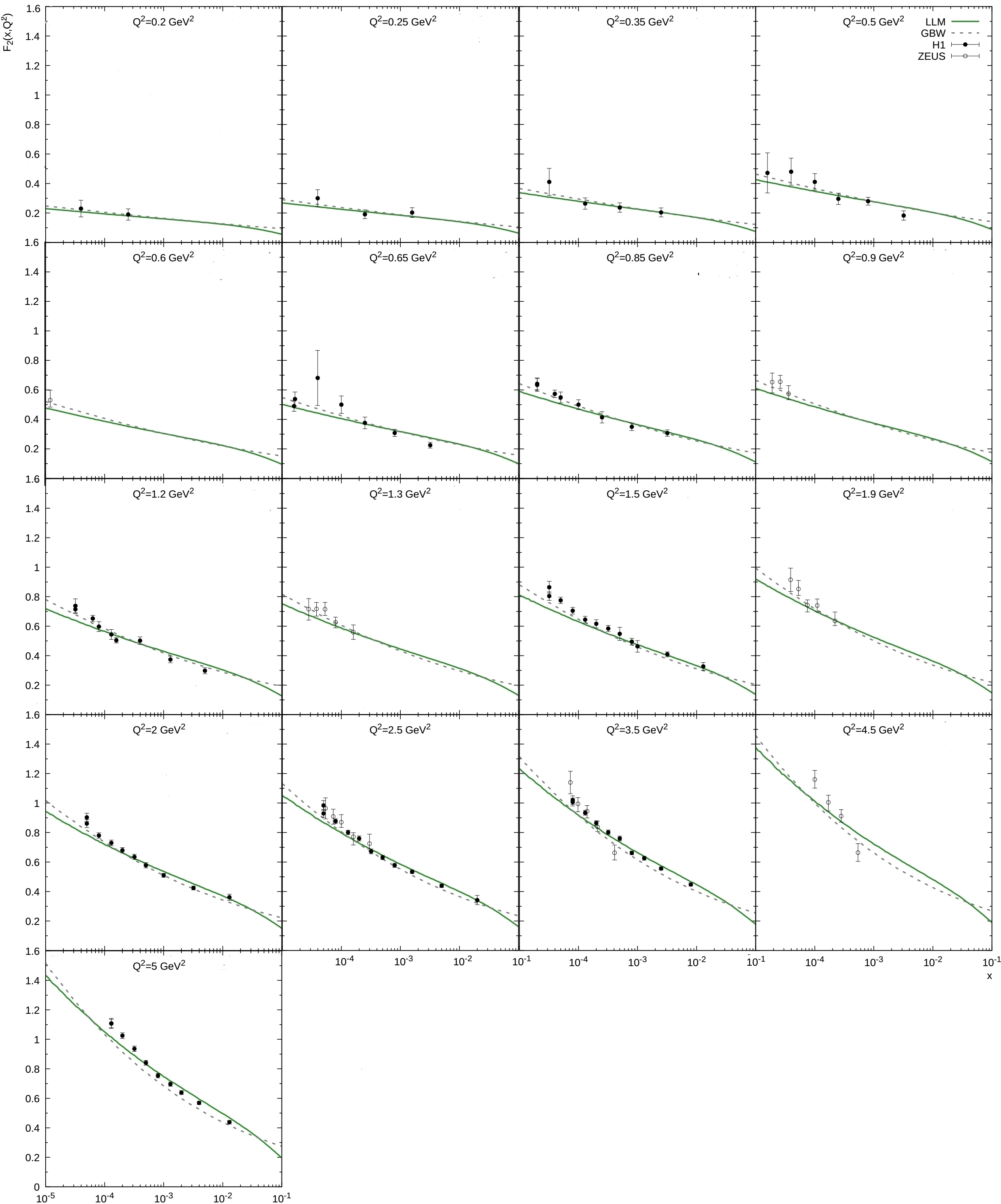}
  \caption{The structure function $F_2$ as a function of $x$ at different $Q^2$. The notations of the lines are the same as on Fig.~\ref{sigred1}.
      The data are taken from ZEUS~\cite{sigma-gamma-proton-ZEUS} (white circles) and H1~\cite{sigma-gamma-proton-H1} (black circles).
}
\label{F2-res}
\end{center}
\end{figure}

Having determined the parameters we can now easily calculate the structure function $F_2(x,Q^2)$. We show our results in comparison with low $Q^2$ ZEUS~\cite{sigma-gamma-proton-ZEUS} and H1~\cite{sigma-gamma-proton-H1} data in Fig.~\ref{F2-res}. A good agreement of LLM results with data is achieved.

%that in order to obtain describe $\sigma_{\gamma^* p}$ and $F_2(x,Q^2)$
%within our approach we
%changed the parameters $x_0$ and $Q_0$ and found them from the best description of
%these parameters. In Fig.~(\ref{fig1}) one can see from this figure that
%$x_0=$ 2.2e-11 and $Q_0=$ 2.2(GeV$/$c)$^2$
%In Fig.~\ref{fig1} the effective cross section
%$\sigma_{\gamma^* }=\sigma_T+\sigma_L$  of the $\gamma^* p$ interaction
%calculated using Eq.~(\ref{def:GBW_sigTL}) and the gluon TMDs
%of Eq.~(\ref{eq:OurGluon}) is presented as a function
%of $Q^2$ at different $W$ and compared with the ZEUS and H1 data
%\cite{ZEUS:2001,H1:2009}. One can see the satisfactory description of
%data at $Q^2\leq$ 4 (GeV$/$c)$^2$ . Let us note that we analyze data at $\mu$
%in the nonperturbatie region, i.e., below $\mu_0\simeq$2-3 GeV, when the
%CCFM evolution equation can start \cite{CCFM}.
 
\begin{figure}
\begin{center}
\includegraphics[width=15cm]{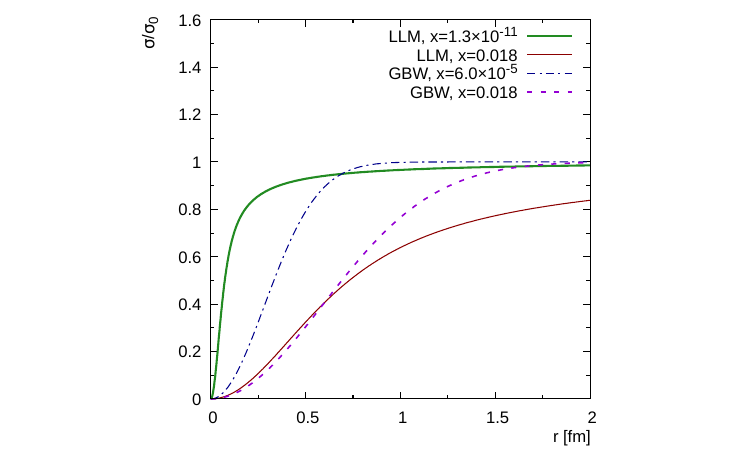}
%sigma-test-x0-allW.pdf
% {FL-ZEUS.pdf}
\caption{The effective dipole cross section $\hat \sigma(x, r^2)$
    normalized by $\sigma_0$ 
  calculated as a function of $r$ at different values of $x$.
}
\label{fig2}
%fig:sigdip_r
\end{center}
\end{figure}

In Fig.~\ref{fig2} we plot the effective dipole cross section
  $\hat \sigma(x, r^2)$ evaluated 
according to~(\ref{def:GBW_dpcrsect})
as a function of $r$ at different values
of $x$. We find that its saturation dynamics at large $r$ strongly 
depends on $x$ and TMD gluon density in a proton. 
In accordance with~(\ref{def:GBW_crsect}),
the GBW gluon density results in the saturation in the region of $r_s\sim 2/R_0$
(see also discussion\cite{GBW2}). 
The corresponding saturation scale at $x = x_0 = 4.2 \cdot 10^{-5}$ and $Q_0= 1$~GeV 
is $Q_s\simeq 2/r_s\simeq 0.8$~GeV.
The LLM gluon TMD results in approximately the same $Q_s$ at fitted
$x_0 = 1.3 \cdot 10^{-11}$ 
and $Q_0 = 2.2$~GeV, see Fig.~\ref{fig2}.
The effective dipole cross section,
${\tilde\sigma}^{\gamma^* p}(r, Q^2)$ and  
gluon density $f_g(x, {\mathbf k}_T^2)$
do not depend on $Q^2$
at low $Q^2 < Q_s^2$, as it shown in \cite{GBW1,LLM-2022}.

%In Fig.~\ref{fig3} the $F_2(x,Q^2)$ at different low $x$ and
%$Q^2\leq $ 4 (GeV$/$c)$^2$ are presented.

%%%%%%%%%%%%%%%%%%%%%%%%%%%%%%%%%%%%%%%%%%%%%%%%%%%%%%%%%%%%%%%%%%%%%%%%%%

%\begin{figure}
%\begin{center}
%\includegraphics[width=5cm]{F2_Q2=1p5.png}
%\hspace*{2cm}
%\includegraphics[width=5cm]{F2_Q2=1p9.png}
%\includegraphics[width=5cm]{F2_Q2=2.png}
%\hspace*{2cm}
%\includegraphics[width=5cm]{F2_Q2=2p5.png}
%\caption{The proton structure function $F_2(x,Q^2)$
%  at different low $x$ and $Q^2\leq$ 4 (GeV$/$c)$^2$ compared to
%  ZEUS \cite{ZEUS:2001}  H1 \cite{H1:2009}
%  and H1 \cite{H1:2009} data.}
%\label{fig3}
%\end{center}
%\end{figure}
%%%%%%%%%%%%%%%%%%%%%%%%%%%%%%%%%%%%%%%%%%%%%%%%%%%%%%%%%%%%%%%%%%%%%%%%%%%%%%

\begin{figure}
\begin{center}
  \includegraphics[width=15cm]{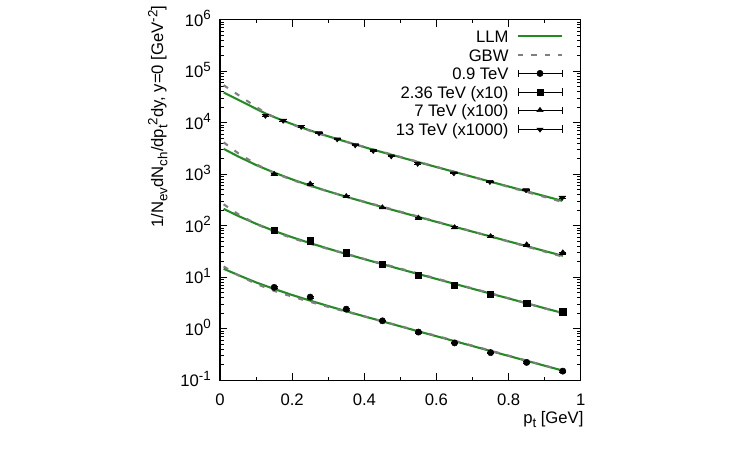}
  \caption{Transverse momentum distributions of soft charged
      hadrons produced in $pp$ collisions at different LHC energies
      in the mid-rapidity region. The notations of the lines are the same as in Fig.~\ref{sigred1}. The experimental data are from
      \cite{SoftHadronData-1, SoftHadronData-2, SoftHadronData-3}.}
\label{fig4}
%fig:sigdip_r
\end{center}
\end{figure}

%%%%%%%%%%%%%%%%%%%%%%%%%%%%%%%%%%%%%%%%%%%%%%%%%%%%%%%%%%%%%%%%%%%%%%%%%%%%%%
Now we turn to the production of soft charged hadrons in $pp$
  collisions
at the LHC energies. These processes are very sensitive to the gluon 
density in a proton at low scales $\mu \sim p_T \simeq 1$~GeV.
%and some parameters of the LLM gluon have been determined from corresponding
%experimental data\cite{?}.
In the calculations we employ the modified QGSM
\cite{ModifiedSoftQuarkGluonStringModel-1, ModifiedSoftQuarkGluonStringModel-2}
and strictly follow the approach described earlier\cite{LLM-2022}.
So, the inclusive hadron spectrum at low $p_T$
and mid-rapidities splits into two pieces: the quark contribution 
$\rho_q$ and the gluon one $\rho_g$:
\begin{gather}
  \rho(x, p_T)= E {d^3\sigma \over d^3 p} \equiv {1\over \pi}
      {d^3 \sigma \over d^2 p_T dy} = \rho_q(x,p_T) + \rho_g(x,p_T).
\end{gather}
\noindent
First term is calculated within the conventional QGSM
\cite{SoftQuarkGluonStringModel-1, SoftQuarkGluonStringModel-2}
using only one-Pomeron exchange. It can be done because in the mid-rapidity and
small $x_T=2p_T/\sqrt{s}$ the multi-Pomeron exchanges give negligibly small
contributions\cite{ModifiedSoftQuarkGluonStringModel-2}. 
The second one, $\rho_q(x,p_T)$, is calculated
as a convolution of the "modified" gluon distribution with 
fragmentation function (FF) of gluons into hadrons
$G_{g\rightarrow h}(z, |{\mathbf {\tilde p}}_T|)$ multiplied by the
inelastic $pp$ cross section (see, for example,\cite{LLM-2022}
  for more details). 
These FFs were calculated at leading (LO) and next-to-leading (NLO) orders\cite{FFs}
and presented in a factorized form
$G_{g\rightarrow h}(z, |{\mathbf {\tilde p}}_T|) = G_{g\rightarrow h}(z) I_h^g(|{\mathbf {\tilde p}}_T|)$, where $I_h^g(|{\mathbf {\tilde p}}_T|)$
could be approximated as:
\begin{gather}
  I_h^g(|{\mathbf {\tilde p}}_T|)=\frac{B_h^g}{2\pi}\exp(-B_h^g
  |{\mathbf {\tilde p}}_T|).
  \label{def:FF_T}
\end{gather}
\noindent
Here $\tilde p_T = p_T - z k_T$ with $p_T$ and $k_T$
being the transverse momenta of the produced hadron $h$ and gluon, respectively.
The FFs of quarks and/or diquarks into hadrons $h$ can be 
expressed in a similar way. 
Below we repeat the calculations\cite{LLM-2022}
with the LLM gluon density and newly fitted values of $x_0$,
$\lambda$ and $Q_0 = 2.2$~GeV.
We find that the best description of the LHC data
\cite{SoftHadronData-1, SoftHadronData-2, SoftHadronData-3} collected at
different energies ($\sqrt s = 0.9$, $2.36$, $7$ and $13$~TeV)
is achieved with slope parameters $B_h^g = 4.35$~GeV$^{-1}$,
$B_h^q = B_h^{qq} = 5.42$~GeV$^{-1}$ ($\chi^2/n.d.f.=3.5$).
All other parameters involved into the calculations and listed in~\cite{LLM-2022}
are unchanged.
Our results are shown in Fig.~\ref{fig4}.
One can see that good agreement with the measurements is
obtained in a wide range of energies. For a comparison we also show results obtained with the GBW TMD. It results in a different fit of the fragmentation parameters $B_h^g = 4.75$~GeV$^{-1}$,
$B_h^q = B_h^{qq} = 6.5$~GeV$^{-1}$, which, however, gives a worse description ($\chi^2/n.d.f.=5.9$).
So, it seems that the proposed expression~(\ref{eq:OurGluon}) leads to better 
description of the considered experimental data, than the GBW model.

Finally, we have verified the sensitivity of inclusive DIS observables
  to the parameters
of the color dipole model, namely, $x_0$, $\lambda$ and the initial scale $Q_0$. 
The dipole cross section ${\hat\sigma}(x, r^2)$ is practically not sensitive
to these parameters at large $r> 0.5$~fm or $Q^2 < 1$~GeV$^2$.
Then, we have achieved a self-consistent simultaneous description
  of the latest HERA data on
  %$\gamma^* p$
the reduced cross section for the $ep$ DIS and the structure
function $F_2(x,Q^2)$ at low $Q^2\leq$ 4.5 GeV$^2$  
and LHC data on the charged hadron production at small transverse momenta,
$p_T \leq 1$~GeV.
In a forthcoming study, the refined LLM gluon density in a proton 
will be used as the initial condition for the subsequent QCD
evolution and thus extented for any scales.

{\sl Acknowledgements}. We thank S.P.~Baranov, 
H.~Jung, S.~Schmidt  and S.~Taheri Monfared for their interest, important comments and remarks. 
Establishing new TMD and fragentation functions parameters as well as soft hadron production simulations and the dipole cross sections calculations were supported by the 
Russian Science Foundation under grant~22-22-00387. The reduced cross section and structure function calculation and corresponding TMD refitting was performed with support of Russian Science Foundation under grant~22-22-00119.

%%%%%%%%%%%%%%%%%%%%%%%%%%%%%%%%%%%%%%%%%%%%%%%%%%%%%%%%%%%%%%%%%

%\bibliography{F2}

\begin{thebibliography}{10}

\bibitem{LHeC}
LHeC Collaboration and FCC{-}he Study Group{,} J. Phys. G: Nucl. Part. Phys.
  {\bf 48}{,}~110501 (2021).

\bibitem{FCC}
FCC Collaboration{,} Eur. Phys. J. C {\bf 79}{,}~474 (2019).

\bibitem{DGLAP}
V.N. Gribov{,} L.N.~Lipatov{,} Sov. J. Nucl. Phys. {\bf 15}{,} 438 (1972); \\
  L.N.~Lipatov{,} Sov. J. Nucl. Phys. {\bf 20}{,} 94 (1975); \\ G.~Altarelli{,}
  G.~Parisi{,} Nucl. Phys. B {\bf 126}{,} 298 (1977); \\ Yu.L. Dokshitzer{,}
  Sov. Phys. JETP {\bf 46}{,}~641 (1977).

\bibitem{FL-NLO-Thorne-1}
C.D.~White{,} R.S.~Thorne{,} Phys. Rev. D {\bf 74}{,}~014002 (2006).

\bibitem{FL-NLO-Thorne-2}
C.D.~White{,} R.S.~Thorne{,} Phys. Rev. D {\bf 75}{,}~034005 (2007).

\bibitem{HighEnergyFactorization}
S.~Catani{,} M. Ciafaloni{,} F. Hautmann{,} Nucl. Phys. B {\bf 366}{,} 135
  (1991){;} \\ J.C. Collins{,} R.K. Ellis{,} Nucl. Phys. B {\bf 360}{,}~3
  (1991).

\bibitem{kt-factorization}
L.V. Gribov{,} E.M. Levin{,} M.G. Ryskin{,} Phys. Rep. {\bf 100}{,} 1 (1983){;}
  \\ E.M. Levin{,} M.G. Ryskin{,} Yu.M. Shabelsky{,} A.G. Shuvaev{,} Sov. J.
  Nucl. Phys. {\bf 53}{,}~657 (1991).

\bibitem{BFKL}
E.A. Kuraev{,} L.N. Lipatov{,} V.S. Fadin{,} Sov. Phys. JETP {\bf 44}{,} 443
  (1976); \\ E.A. Kuraev{,} L.N. Lipatov{,} V.S. Fadin{,} Sov. Phys. JETP {\bf
  45}{,} 199 (1977); \\ I.I. Balitsky{,} L.N. Lipatov{,} Sov. J. Nucl. Phys.
  {\bf 28}{,}~822 (1978).

\bibitem{CCFM}
M.~Ciafaloni{,} Nucl. Phys. B {\bf 296}{,} 49 (1988); \\ S. Catani{,} F.
  Fiorani{,} G. Marchesini{,} Phys. Lett. B {\bf 234}{,} 339 (1990); \\ S.
  Catani{,} F. Fiorani{,} G. Marchesini{,} Nucl. Phys. B {\bf 336}{,} 18
  (1990); \\ G. Marchesini{,} Nucl. Phys. B {\bf 445}{,}~49 (1995).

\bibitem{Saturation-Mueller}
A.H.~Mueller{,} Nucl. Phys. B {\bf 335}{,}~115 (1990).

\bibitem{DP-NZ-1}
N.~Nikolaev{,} B.G.~Zakharov{,} Z. Phys. C {\bf 49}{,}~607 (1990).

\bibitem{DP-NZ-2}
N.~Nikolaev{,} E.P.~Predazzi{,} B.G.~Zakharov{,} Phys. Lett. D {\bf 326}{,}~161
  (1994).

\bibitem{GBW1}
K.~Golec-Biernat{,} M. W\"usthoff{,} Phys. Rev. D {\bf 59}{,}~014017 (1998).

\bibitem{GBW2}
K.~Golec-Biernat{,} M. W\"usthoff{,} Phys. Rev. D {\bf 60}{,}~114023 (1999).

\bibitem{Input-1}
A.A. Grinyuk{,} A.V. Lipatov{,} G.I. Lykasov{,} N.P. Zotov{,} Phys. Rev. D {\bf
  87}{,}~074017 (2013).

\bibitem{ModifiedSoftQuarkGluonStringModel-1}
V.A. Bednyakov{,} G.I. Lykasov{,} V.V. Lyubushkin{,} Europhys. Lett. {\bf
  92}{,}~31001 (2010).

\bibitem{ModifiedSoftQuarkGluonStringModel-2}
V.A. Bednyakov{,} A.A. Grinyuk{,} G.I. Lykasov{,} M. Poghosyan{,} Int. J. Mod.
  Phys. A {\bf 27}{,}~1250042 (2012).

\bibitem{SoftQuarkGluonStringModel-1}
A.B. Kaidalov{,} Z. Phys. C {\bf 12}{,} 63 (1982){;}\\ A.B. Kaidalov{,} Surveys
  High Energy Phys. {\bf 13}{,} 265 (1999){;}\\ A.B. Kaidalov{,} O.I.
  Piskunova{,} Z. Phys. C {\bf 30}{,}~145 (1986).

\bibitem{SoftQuarkGluonStringModel-2}
G.I. Lykasov{,} M.N. Sergeenko{,} Z. Phys. C {\bf 52}{,} 635 (1991){;}\\ G.I.
  Lykasov{,} M.N. Sergeenko{,} Z. Phys. C {\bf 56}{,} 697 (1992){;}\\ G.I.
  Lykasov{,} M.N. Sergeenko{,} Z. Phys. C {\bf 70}{,}~455 (1996).

\bibitem{Input-2}
A.V. Lipatov{,} G.I. Lykasov{,} N.P. Zotov{,} Phys. Rev. D {\bf 89}{,}~014001
  (2014).

\bibitem{Input-3}
A.A. Grinyuk{,} A.V. Lipatov{,} G.I. Lykasov{,} N.P. Zotov{,} Phys. Rev. D {\bf
  93}{,}~014035 (2016).

\bibitem{Input-4}
N.A. Abdulov{,} H. Jung{,} A.V. Lipatov{,} G.I. Lykasov{,} M.A. Malyshev{,}
  Phys. Rev. D {\bf 98}{,}~054010 (2018).

\bibitem{LLM-2022}
A.V.~Lipatov{,} G.I.~Lykasov{,} M.A.~Malyshev{,} Phys. Rev. D {\bf
  107}{,}~014022 (2023).

\bibitem{sigma_red-ZEUS+H1}
ZEUS and H1~Collaborations{,} Eur. Phys. J. C {\bf 75}{,}~580 (2015).

\bibitem{sigma-gamma-proton-ZEUS}
ZEUS Collaboration{,} Eur. Phys. J. C {\bf 7}{,}~609 (1999).

\bibitem{sigma-gamma-proton-H1}
H1~Collaboration{,} Eur. Phys. J. C {\bf 63}{,}~625 (2009).

\bibitem{SoftHadronData-1}
ATLAS Collaboration{,} New J. Phys. {\bf 13}{,}~053033 (2011).

\bibitem{SoftHadronData-2}
CMS Collaboration{,} Phys. Rev. Lett. {\bf 105}{,}~022002 (2010).

\bibitem{SoftHadronData-3}
ATLAS Collaboration{,} Eur. Phys. J. C {\bf 76}{,}~502 (2016).

\bibitem{TMDLib2}
N.A. Abdulov{,} A. Bacchetta{,} S.P. Baranov{,} A. Bermudez Martinez{,} V.
  Bertone{,} C. Bissolotti{,} V. Candelise{,} L.I. Estevez Banos{,} M. Bury{,}
  P.L.S. Connor{,} L. Favart{,} F. Guzman{,} F. Hautmann{,} M. Hentschinski{,}
  H. Jung{,} L. Keersmaekers{,} A.V. Kotikov{,} A. Kusina{,} K. Kutak{,} A.
  Lelek{,} J. Lidrych{,} A.V. Lipatov{,} G.I. Lykasov{,} M.A. Malyshev{,} M.
  Mendizabal{,} S. Prestel{,} S. Sadeghi Barzani{,} S. Sapeta{,} M. Schmitz{,}
  A. Signori{,} G. Sorrentino{,} S. Taheri Monfared{,}~A. van Hameren{,}
  A.M.~van Kampen{,} M. Vanden Bemden{,} A. Vladimirov{,} Q. Wang{,} H. Yang{,}
  Eur. Phys. J. C {\bf 81}{,} 752~(2021).

\bibitem{FL-our}
A.V.~Lipatov{,} G.I.~Lykasov{,} M.A.~Malyshev{,} Phys. Lett. B {\bf
  839}{,}~137780 (2023).

\bibitem{Photon-our}
A.V.~Lipatov{,} M.A.~Malyshev{,} Phys. Rev. D {\bf 108}{,}~014025 (2023).

\bibitem{Kutak1}
T.~Goda{,} K. Kutak{,} S. Sapeta{,} Nucl. Phys. B {\bf 990}{,}~116155 (2023).

\bibitem{Kutak2}
T.~Goda{,} K. Kutak{,} S.~Sapeta{,} arXiv:2305.14025~[hep ph].

\bibitem{FFs}
J.~Binnewies{,} B.A. Kniehl{,} G. Kramer{,} Phys. Rev. D {\bf 52}{,}~4947
  (1995).

\end{thebibliography}

\end{document}